\documentclass{chjaa}               % preprint, the final version for publication
\usepackage{graphicx,times}             %for PS/EPS graphics inclusion, new
\input{epsf.sty}                        %for PS/EPS graphics inclusion, old
\input{psfig.sty}                       %for PS/EPS graphics inclusion, old

\begin{document}

   \title{X-ray and optical followup of gamma-ray (up to TeV) sources}
%\footnotetext{$*$ Supported by the National Natural Science Foundation of China.}}
%   \subtitle{I. Place Your Subtitle Here}

   \volnopage{Vol.0 (200x) No.0, 000--000}      %%preserved for Editor. DOn't remove!
   \setcounter{page}{1}           %%starting page, preserved for Editor. DOn't remove!

   \author{Nicola Masetti
      \inst{1}\mailto{masetti@iasfbo.inaf.it}
%% Please move "\mailto{}" to the corresponding author of the paper.
       }

   \institute{INAF--Istituto di Astrofisica Spaziale e Fisica Cosmica di Bologna, 
             via Gobetti 101 -- I-40129 Bologna, Italy\\
             \email{masetti@iasfbo.inaf.it}
%% offprint requests will be sent. Note to pair \mailto{} with \email{}
          }

   \date{Received~~2007 June 30; accepted~~}

   \abstract{The imaging capabilities of the {\it INTEGRAL} and {\it HESS} observatories allow
the study of hard X--ray and TeV sources with unprecedented positional accuracy. Here I
review the multiwavelength followup studies which are currently being performed on the 
unidentified sources detected by these facilities in order to unveil their actual nature.
   \keywords{X--rays: general --- Gamma rays: observations --- Astrometry --- 
Techniques: spectroscopic --- Methods: observational}
   }

   \authorrunning{N. Masetti}            %author_head in even pages
   \titlerunning{Followup of gamma-ray (up to TeV) sources}  % title_head in odd pages

   \maketitle
%% The author head (on even pages) and the title head (on odd pages) will be
%% automatically extracted from \author{} and \title{}. Whenever the title is too long,
%% you will be asked to supply a shorter one by inserting either \authorrunning{} or
%% \titlerunning{} before \maketitle. Anyway, you can specify your own heads in advance.
%%
%%
%% Note: In the following text body of your manuscript, please note several differences from
%%       other major journals:
%% (1) \subsection{Please Capitalize the First Letter of Each Notional Word in Subsection Title}
%% (2) Please Capitalize the First Letter of Each Notional Word in all tables' captions

%
%________________________________________________ sections below
%
\section{Introduction}           %% first-level sections will be auto-capitalized

The last years have witnessed a major advance in the study of hard X--ray 
and TeV emitting cosmic objects, particularly for what concerns the 
accuracy of their localization. This accomplishment was mainly achieved 
thanks to two observatories: the {\it INTEGRAL} satellite (Winkler et al. 
2003) for hard X--ray sources and the {\it HESS} \v{C}erenkov telescope 
array (Hinton 2004) for TeV objects. Both facilities are indeed capable to 
localize high-energy objects with a precision of a few arcminutes. This 
possibility allows reducing the area of the sky in which one should search 
for the counterpart of these objects at longer wavelengths in order to 
study their broadband spectrum and, eventually, to understand their actual 
nature.

This contribution thus aims at reviewing the current knowledge and results 
of this search for lower-energy counterparts (mainly at soft X-ray and 
optical frequencies) of these unidentified {\it INTEGRAL} and {\it HESS} 
sources.

\section{The search for counterparts to {\it INTEGRAL} sources}

This section may be considered as an update of my contribution at
the proceedings of the Frascati Workshop held 2 years ago in Vulcano
(Masetti 2006).

Recently, the largest survey performed with the IBIS instrument onboard
{\it INTEGRAL} has been issued (Bird et al. 2007). It spans nearly 4
years of observations (from October 2002 to April 2006), for a total
time of about 40 Ms. This resulted in a catalogue (Fig. 1) with an average 
sensitivity limit of $\sim$1 mCrab (depending on the observing time 
spent on a given zone of the sky) in the 20--100 keV band and which contains 
421 sources with positional accuracy spanning from 1 to 5 arcmin, depending 
on the source intensity. Of these, 147 (35\%) are X--ray Binaries, 118 (28\%) 
are Active Galactic Nuclei (AGNs), 23 (5\%) are Cataclysmic Variables (CVs);
among the remaining sources, 115 of them (about 27\% of the total) have 
unidentified nature.

\begin{figure}
\begin{center}
\hspace{.5cm}
\psfig{figure=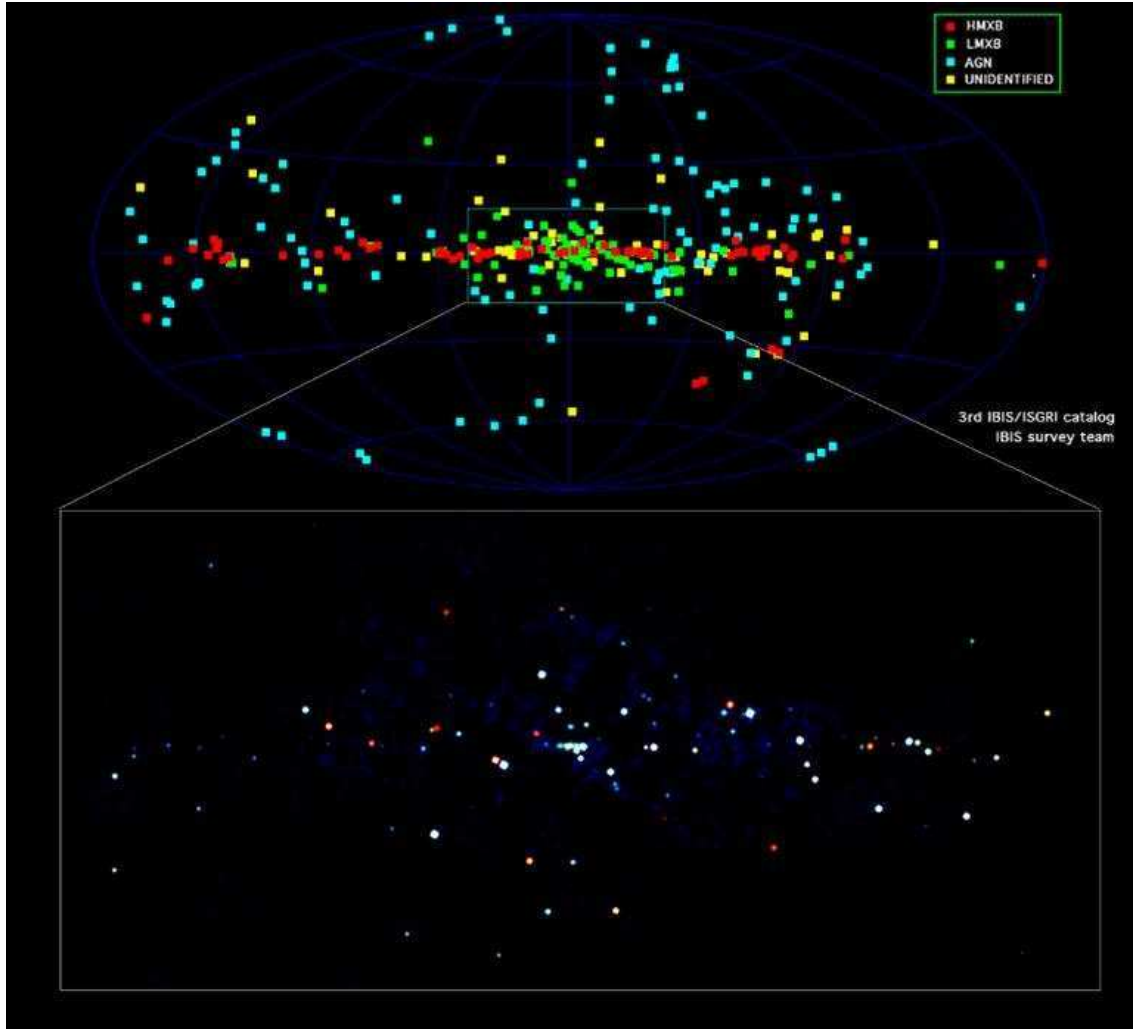,width=150mm,angle=0.0}
\caption{{\it Upper image}: distribution on the sky of three of the main 
soft gamma-ray source populations (HMXBs, LMXBs and AGNs) observed in the 
third {\it INTEGRAL}/IBIS survey catalogue (Bird et al. 2007). 
Besides, around one out of four of the sources seen by 
{\it INTEGRAL} are unidentified, and their distribution is also shown. 
{\it  Lower image}: false colour image of the central region of our Galaxy. 
This is a composite image based on all-sky IBIS/ISGRI maps in three energy 
windows between 17 and 100 keV and represents the true X--ray `colours' of 
the sources. Red sources are dominated by emission below 30 keV, while blue 
sources have harder spectra, emitting strongly above 40 keV (Credit: IBIS 
Survey Team; {\it INTEGRAL} Picture of the Month of February 2007).}
\end{center}
\end{figure}

Besides this one, a number of other surveys were published recently;
in particular, the one of Krivonos et al. (2007), of comparable depth
with respect to that of Bird et al. (2007), detected 400 sources with a
similar fraction of unidentified objects. In parallel, Bodaghee et al. 
(2007) collected all the available information on the sources detected 
by {\it INTEGRAL} during the first 4 years of operations, thus producing a 
catalogue of about 500 sources: again, the percentage of unclassified 
sources (26\%) is similar to that found by Bird et al. (2007).

As already widely discussed in the past, the {\it INTEGRAL} unprecedented
capabilities of localizing hard X--ray sources, together with the 
positional cross-correlation with catalogues at longer wavelengths (especially
the ones of soft X--ray satellites), allows a further reduction of the error
box to few arcseconds or less. Indeed, Stephen et al. (2006) demonstrated that,
when a soft X--ray source is found within the IBIS error box, it is most likely
the counterpart of the {\it INTEGRAL} object.
This technique allows pinpointing the putative optical and near-infrared (NIR) 
counterparts of the {\it INTEGRAL} sources, on which spectroscopy can be
performed to ascertain the actual nature of the high-energy emitting object.

This approach was first applied, for a newly-discovered {\it INTEGRAL} source,
IGR J16138$-$4848, by Filliatre \& Chaty (2004): thanks to the arcsec-sized
X--ray position afforded with {\it XMM-Newton}, they could identify the
nature of this source as a heavily absorbed High-Mass X--ray Binary (HMXB) 
hosting a supergiant B[e] star.

Motivated by this, on July 2004 our group started a program for the 
systematic identification of {\it INTEGRAL} sources of unknown nature 
through optical spectroscopic observations of putative optical candidates 
pinpointed through the use of soft X--ray observations of {\it INTEGRAL} 
error boxes performed with {\it ROSAT}, {\it Chandra}, {\it XMM-Newton} or 
{\it Swift}. This program, performed at northern (Loiano, Asiago) and 
southern (ESO, SAAO, CTIO) telescopes, allowed us to identify 40 objects, 
of which 23 are nearby AGNs (with redshift between 0.013 and 0.084: among 
them we found 11 Seyfert 1 galaxies, 11 Seyfert 2 galaxies, and 1 BL Lac), 
10 are HMRBs and 7 are CVs (Masetti et al. 2004, 2006a,b,c,d).

Using cross-correlations with available catalogues of variable or 
emission-line stars, we could also identify 3 more {\it INTEGRAL} sources 
(1 transient HMXB and 2 Symbiotic stars: see Leyder et al. 2007 and 
Masetti et al. 2005, 2006e,f). These three identifications were confirmed 
by X--ray observations made with {\it Swift} (Tueller et al., 2005a,b; 
Kennea \& Campana 2006).

Since the second part of 2006 this observational campaign was continued at 
Loiano, Asiago, CTIO, CASLEO and SAAO observatories, and more observations 
were collected at CTIO and ESO in June 2007. Besides, we retrieved 
archival spectra of putative optical counterparts from the 6dF (Jones et 
al. 2004) and SDSS (Adelman-McCarthy et al. 2006) surveys. Preliminary 
results (Masetti et al. 2006g,h, 2007a,b) from these observations showed 
that, of 15 newly identified objects, 13 are AGNs (with 0.006 $<z<$ 0.230: 
among them, 4 Seyfert 1s and 9 Seyfert 2s) and 2 are Low-Mass X--ray 
Binaries (LMXBs) or reddened CVs.

In particular, the most recent (April-May 2007) observations performed at 
SAAO showed that (i) the galaxy associated with IGR J16024$-$6107 is at 
$z$=0.011 and its emission line ratios place it at the border between 
Seyfert 2 AGNs and starbursts/HII galaxies; (ii) IGR J14298$-$6715 is a 
Galactic emission-line object, most likely a LMXB or a CV; (iii) IGR 
J19405$-$3015 is a Seyfert 1.2 galaxy at $z$=0.052.

The need for high-precision localizations (at the arcsecond level or 
better) assumes fundamental importance when we are dealing with crowded 
optical fields, for instance in the case of searches along the Galactic 
Plane. In this sense the case of the transient X--ray source IGR 
J17497$-$2821 is paradigmatic: the {\it Swift}/XRT position, with an error 
radius of 5$''$, contains more than 20 NIR sources, and only the precise 
localization (better than 1$''$) afforded by {\it Chandra} allowed 
determining the actual counterpart of this X--ray source (Torres et al. 
2006; Paizis et al. 2007).

\begin{figure}
%\begin{center}
%\vspace{-3cm}
\psfig{figure=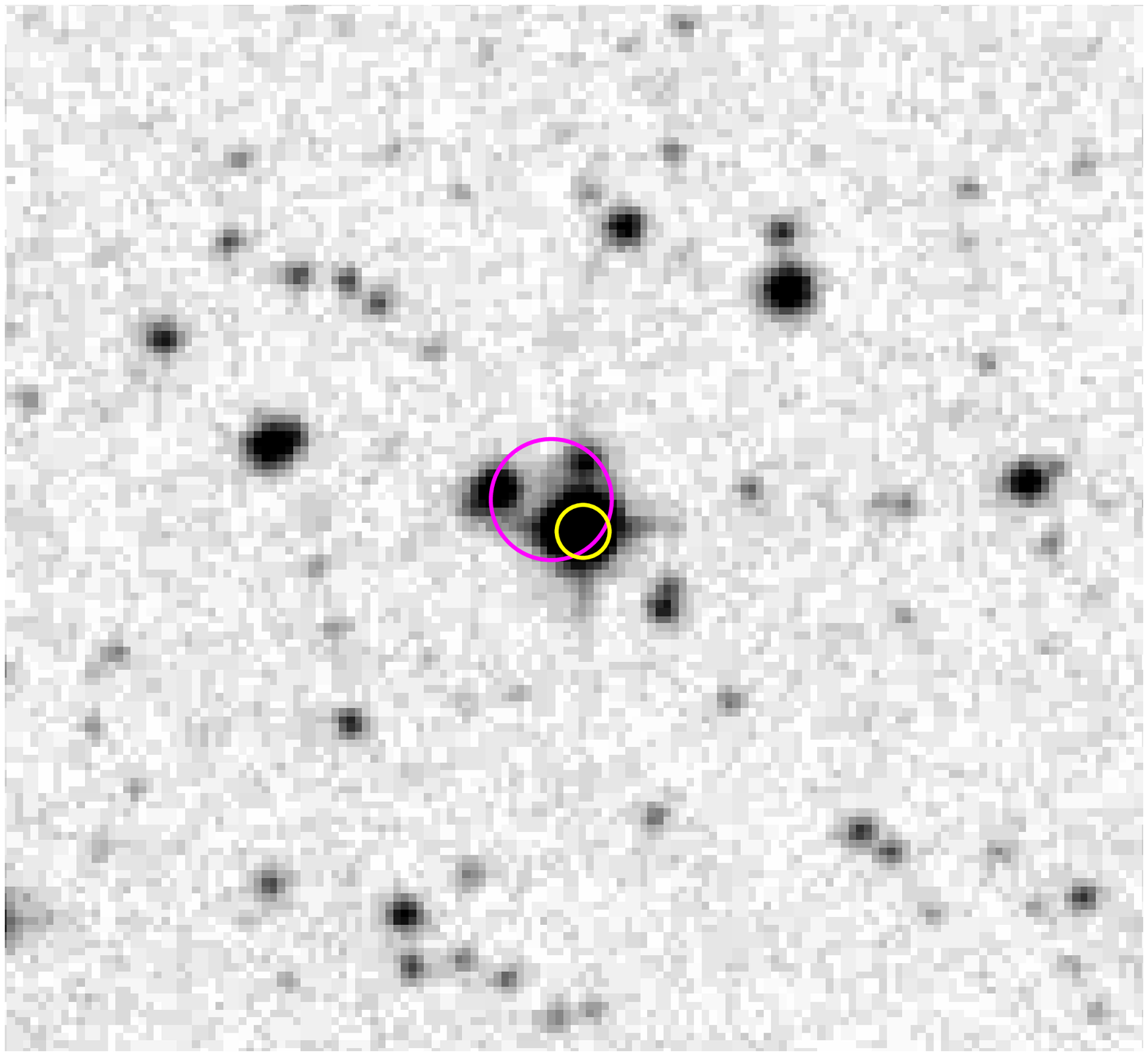,width=55mm,angle=0.0}

\vspace{-8cm}
\hspace{5cm}
\psfig{figure=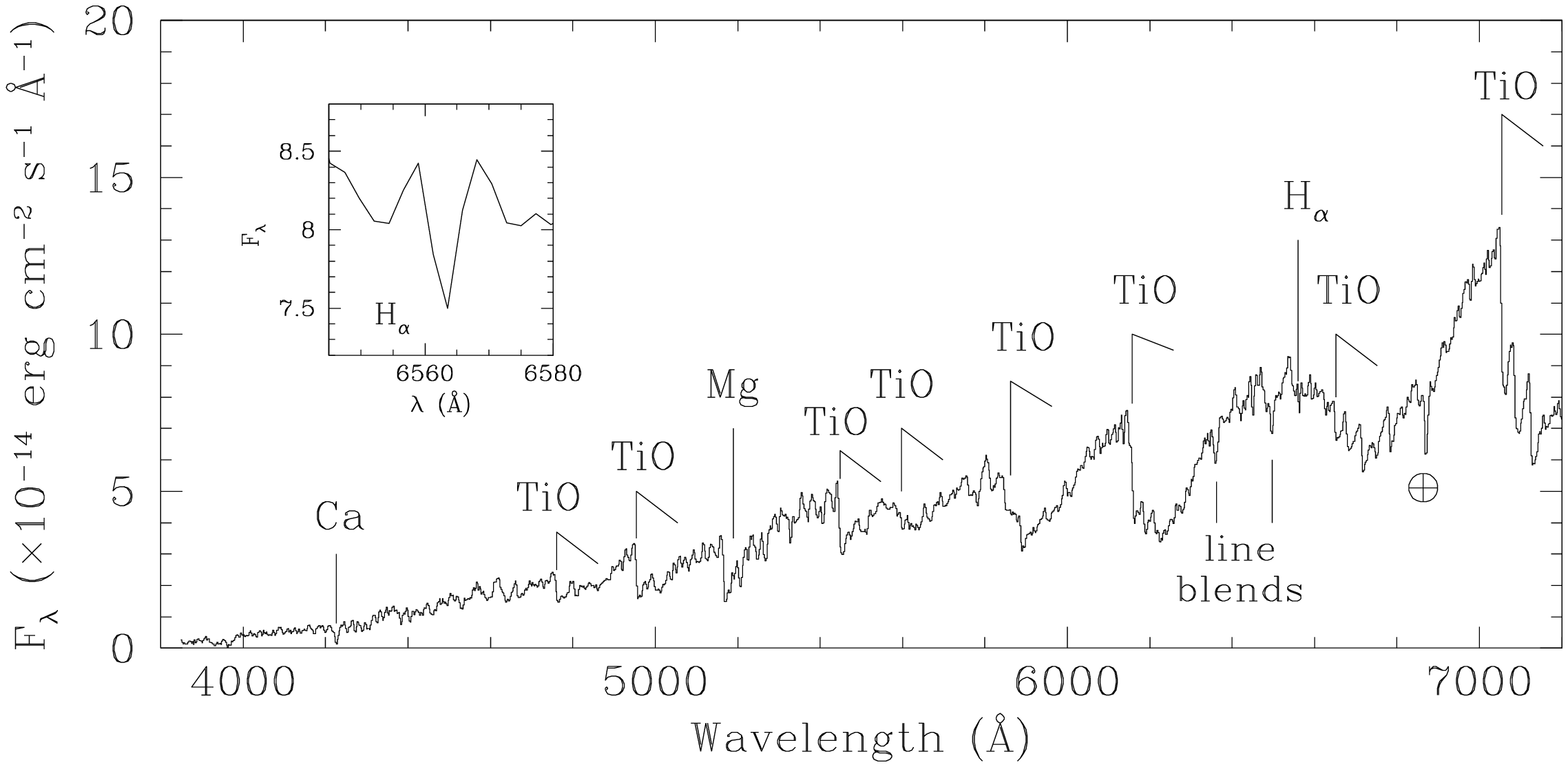,height=85mm,angle=-90.0}
\caption{{\it Left panel}: optical image of the field of IGR J16194$-$2810 
with the {\it Swift}/XRT (smaller circle, 3$\farcs$5 radius) and the 
{\it ROSAT}/PSPC (larger circle, 8$''$ radius) 
X--ray positions superimposed. The only star positionally consistent with the 
XRT error circle is the brighter one at the centre of the image.
In the figure, North is at top and East is to the left. The field size is 
$\sim$2$\farcm$5$\times$2$\farcm$5. {\it Right panel}: optical spectrum of 
the counterpart of IGR J16194$-$2810 obtained with the 1.9-meter Radcliffe 
telescope of SAAO (South Africa). The spectrum is typical of a star of type 
M2\,III. The inset shows a close-up of the spectrum around the H$_\alpha$ 
region. See Masetti et al. (2007c) for details.}
%\end{center}
\end{figure}

Because of the paramount importance of exploring the {\it INTEGRAL} error 
circle of the unidentified sources with soft X--ray satellites allowing 
arcsecond-sized precision on the object position, we started a 
collaboration with the {\it Swift} team to get short ($\sim$5 ks) 
exposures on the error boxes of {\it INTEGRAL} sources lacking soft X--ray 
pointings or with soft X--ray positions with radius of tens of arseconds 
and containing several optical sources. This was done in order to get a 
position as much precise as possible for these unidentified high-energy 
sources.

This collaboration has proven very fruitful, and the preliminary results 
are more than encouraging. Among them, the XRT observations of the {\it 
INTEGRAL} source IGR J16194$-$2810 allowed us to spot the soft X-ray 
counterpart and to see that it behaves as an X-ray binary under both the 
spectroscopic and temporal profiles. These X--ray pointings also permitted 
us to pinpoint the optical counterpart (Fig. 2, left panel): optical 
spectroscopy revealed that it is a `'normal' red giant of spectral type 
M2 III, with no emission features (Fig. 2, right panel). All this joint 
multiwavelength information allowed us to prove that this {\it INTEGRAL}
source is a Symbiotic X-ray Binary (SyXB; see Masetti et al. 2007c for 
details), i.e. a LMXB composed of a compact object (likely a neutron star) 
accreting from the wind of a red giant. We stress that this object is 
remarkable in the sense that SyXBs are very rare LMXBs: IGR J16194$-$2810 
is just the fifth case known to date. The other four known cases are
GX 1+4 (Chakrabarty \& Roche 1997), 4U 1700+24 (Garcia et al. 1983; 
Masetti et al. 2002, 2006i), 4U 1954+31 (Masetti et al. 2006i, 2007d) and
Sct X-1 (Kaplan et al. 2007).

If we now turn to analyze the sample of {\it INTEGRAL} objects in the 
three IBIS surveys (see Fig. 3, from Bird et al. 2007) and we consider 
their breakdown into the various classes, we can see the following when we 
pass from the 1$^{\rm st}$ to the 3$^{\rm rd}$ survey catalogue:

\begin{itemize}

\item
a large increase of AGN detections; 

\item
a substantial increase of CV detections;

\item
the percentage of unidentified sources keeps nearly constant (22\%, 27\% 
and 27\% in the three catalogues, respectively);

\item
a decrease in the detection rate of LMXBs.

\end{itemize}

\begin{figure}
\hspace{1.5cm}
\psfig{figure=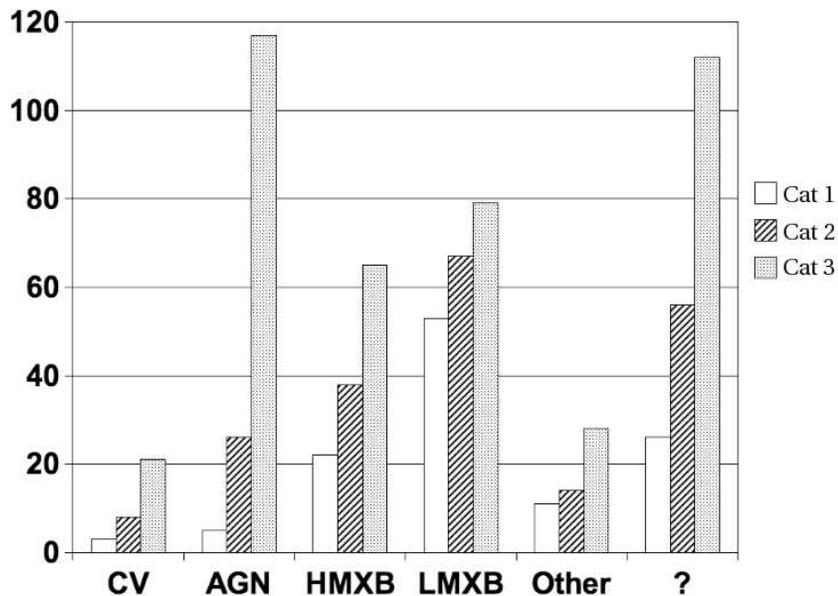,width=120mm,angle=-90.0}
\caption{Histogram reporting the numbers of sources in the first, second, and 
third IBIS/ISGRI catalogs, classified by type; the question mark indicates
the unidentified objects (from Bird et al. 2007).}
\end{figure}

The first item is reasonably explained by the fact that {\it INTEGRAL} 
allows the exploration of the region around the Galactic Plane (the 
so-called `Zone of Avoidance' for extragalactic studies) at hard X--rays, 
which can easily penetrate the dust layers of this part of the sky 
bringing information on what lies behind it.

According to Bodaghee et al. (2007), the last item is explained by the 
fact that LMXBs are intrinsically less obscured than HMXBs and AGNs, and 
were therefore easier to detect with previous satellites. Besides, the 
unprecedented localization capabilities of {\it INTEGRAL} in the energy 
range above 20 keV and its high sensitivity in this band are two further 
reasons which allow this satellite to detect many more (obscured) HMXBs 
than LMXBs.

The detection of a non-negligible fraction (about 5\% of the 3$^{\rm rd}$ 
{\it INTEGRAL} survey sources, and $\sim$10\% of the optical 
identifications achieved up to now) of CVs is instead surprising, as they 
were not expected to substantially emit in the hard X--ray band. As a 
matter of facts, it is found that most of these are magnetic CVs 
(intermediate polars or polars: Barlow et al. 2006; Masetti et al. 2006d), 
which may indeed show hard X--ray tails (e.g., de Martino et al. 2004).

Besides all this, {\it INTEGRAL} allowed the detection of a new class of
heavily absorbed Galactic HMXBs (see the contribution by S. Chaty to 
these proceedings).

Thus, the synergy among {\it INTEGRAL}, soft X--ray satellites like {\it 
Swift}, and optical spectroscopy is allowing the discovery of new classes 
of sources, of new members of rare classes of objects, or of unexpected 
features from known types of objects. It is thus planned to extend this 
very effective approach of following up the error boxes of unidentified 
{\it INTEGRAL} sources at soft X--rays through the use of the X--ray 
satellite {\it XMM-Newton}, in order to get more high-precision positions 
in the near future.

To conclude this Section, I would like to mention the web page by Jerome 
Rodriguez which contains an updated list of detected {\it INTEGRAL} 
sources:

\medskip
\noindent
{\tt http://isdcul3.unige.ch/\~{}rodrigue/html/igrsources.html} .

I also report here that, as a service to the community, I am maintaining a 
web archive with the main properties of the {\it INTEGRAL} sources 
identified through optical or NIR spectroscopy. This archive can be found 
at the URL:

\medskip
\noindent
{\tt http://www.iasfbo.inaf.it/IGR/main.html} .

\section{Unidentified {\it HESS} sources and their multiwavelength followup}

Recently, the {\it HESS} team reported the discovery of 14 previously 
unknown TeV gamma-ray sources (Aharonian et al. 2006); some of these 
detections have subsequently been confirmed by deeper {\it HESS} 
observations or by pointings made with the {\it MAGIC} \v{C}erenkov 
telescope.

The common properties of these sources are the following: (i) they are 
positioned along the Galactic Plane; (ii) many of them appear to be 
extended; (iii) their energy spectra are generally hard, with an average 
photon index $\Gamma \sim$ 2.3.

Some of them were readily identified as SuperNova Remnants (SNRs) or 
Pulsar Wind Nebulae (PWNe) through positional coincidence; however, many 
still lack counterparts at longer wavelengths, and hence a clear 
identification of their nature. Aharonian et al. (2006) also put forward 
the hypothesis that those of them lacking longer-wavelength counterpart 
may be classified as `'dark particle accelerators', i.e. a new class of 
sources which are relatively bright at TeV energies and much fainter in 
the other bands of the electromagnetic spectrum. Nonetheless, other types 
of sources can emit TeV photons, such as SNRs, pulsars and PWNe, 
microquasars (or, more generally, X--ray binaries), and background AGNs.

Given the great importance of these accurate detections and the potential
discovery of a new class of sources, the search of counterparts at other
wavebands was soon initiated by several groups around the world.

The first step in this search was to find positional agreement between the 
various {\it HESS} sources and the object in catalogues at other 
wavelengths, mostly in the soft X--ray and radio bands. In parallel, 
followup programs at longer wavelengths were started, mostly with the use 
of X--ray satellites, such as {\it Chandra}, {\it XMM-Newton}, {\it Swift} 
and {\it INTEGRAL} itself.

After this, and once a possible longer-wavelength counterpart was found, the 
second step in this work was the search of a viable gamma-ray emission 
mechanism explaining the multiwaveband observations from the putative 
counterpart. Next, the third step was to provide a consistent multi-wavelength 
picture. Additionally, in the cases in which the source is extended, a further
analysis involved the study of the morphological match among the appearance of 
the source itself at various wavelengths.

In what follows I will report the main results of this multiwavelength search
for the counterpart of these {\it HESS} sources at other wavelengths, and I
will focus on the most interesting cases.

\begin{figure}
%\begin{center}
\mbox{\psfig{figure=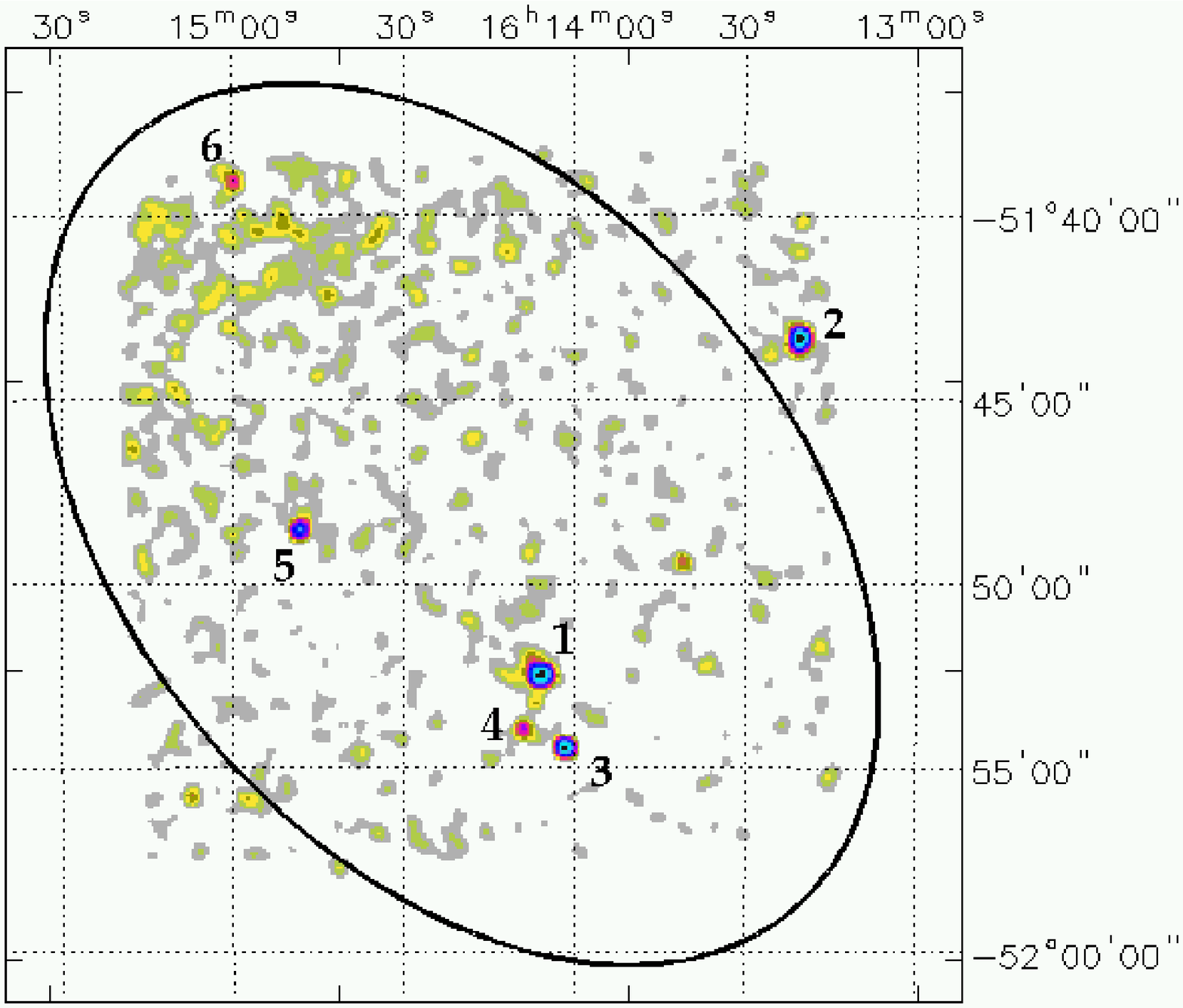,width=75mm,angle=0.0}}
%\vspace{1cm}
\psfig{figure=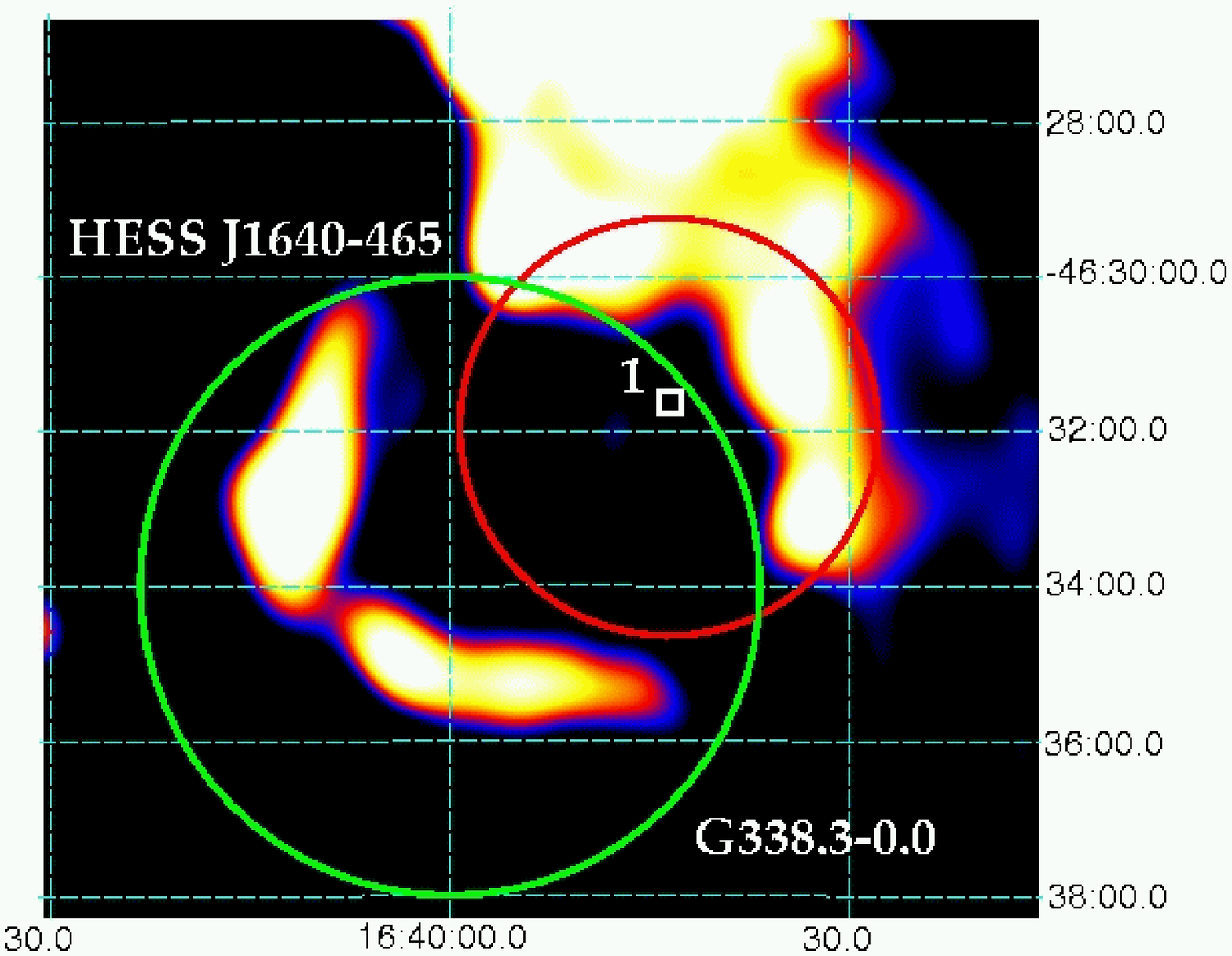,width=85mm,angle=0.0}
\caption{{\it Left panel}: XRT 0.3--10 keV image of the region surrounding 
HESS J1614$-$518. The ellipse (with axes of size 28$'$ and 18$'$) 
represents the extension of the TeV source. 
For details see Landi et al. (2006, 2007a). {\it Right panel}: radio image 
at 843 MHz of the region surrounding HESS J1640$-$465. The larger circle 
(green, 4$'$ radius) describes the position and extension of SNR 
G338.3$-$0.0 as given in Green (2004). The smaller circle (red, 
2$\farcm$5 radius) represents instead the extension of the TeV source. The 
position of the XRT Source 1 is given by a box. For details see Landi et 
al. (2006).}
%\end{center}
\end{figure}

{\bf HESS J1614$-$518}. The {\it Swift}/XRT followup observations on this 
source detected 6 faint X--ray objects (with 2--10 keV flux less than 
5$\times$10$^{-13}$ erg cm$^{-2}$ s$^{-1}$) within or close to the {\it 
HESS} error box (Fig. 4, left panel; Landi et al. 2006, 2007a). Optical 
followup on five of these sources showed that they are Galactic stars 
(Landi et al. 2007a); moreover, the open cluster C1609$-$517, at a distance 
of 1 kpc and with age 40 Myr (Piatti et al. 2000) lies positionally close 
to the XRT source No. 1 and to the centre of the {\it HESS} error ellipse. 
It is possible that at least some of the stars detected in X--rays by {\it 
Swift} are part of this cluster. The recent discovery of the unidentified 
TeV gamma-ray sources in the vicinity of open clusters Cyg OB2 with {\it 
HEGRA} (Aharonian et al. 2002) and Cen OB1 with {\it HESS} (Aharonian et 
al. 2005a) have renewed the interest in young open clusters as possible 
sites for the acceleration of cosmic rays and for the emission of TeV 
gamma-rays; it is thus suggested by Landi et al. (2007a) that the open 
cluster C1609$-$517 is possibly the counterpart of HESS J1614-518.

{\bf HESS J1616$-$508}. According to Matsumoto et al. (2006), {\it Suzaku} 
observations suggest that this source is a Dark Particle Accelerator, as 
no X--ray source was detected within the TeV error circle. The 
brightest high-energy source in the field is PSR J1617$-$5055, a young 
X-ray pulsar already proposed to emit TeV gamma-rays (Torii et al. 1998). 
{\it INTEGRAL}, {\it BeppoSAX} and {\it XMM-Newton} observations (Landi et 
al. 2007b) support this hypothesis, indicating it as the possible 
counterpart of the {\it HESS} TeV source. Indeed, according to Landi et 
al. (2007b), 1.2\% of the spin down luminosity of this pulsar is 
sufficient to power the TeV flux seen by {\it HESS}. However, PSR 
J1617$-$5055 lies slightly outside the {\it HESS} error box.
In its followup observations, {\it Swift} detected 2 X--ray
sources within or close to the {\it HESS} error circle, but they are 
unlikely counterparts of HESS J1616$-$508: one is soft and is probably a 
star, the other is possibly an artifact, (Landi et al. 2007b). 
These authors suggest that an asymmetric PWN powered by PSR J1617$-$5055 
is the counterpart of HESS J1616$-$508; this would explain why the pulsar 
is not exactly within the {\it HESS} error box.

{\bf HESS J1632$-$478 and HESS J1634$-$472}. These two sources are quite 
close to each other (they are separated by about 45$'$) and are 
positionally consistent with the X--ray binaries 4U 1630$-$47, IGR 
J16358$-$4726 and IGR J16320$-$4751: HESS J1634$-$472 with the former two, 
and HESS J1632$-$478 with the latter one. However, in this region of the 
sky (the Norma Arm region) there is a high concentration of sources, so 
the probability of chance coincidence can be high. 
This said, if we consider HESS J1632$-$478, {\it XMM-Newton} archival data 
show that there are actually two more (unidentified) X--ray sources, 
besides IGR J16320$-$4751, within the TeV error ellipse: AX 
J163252$-$4746, at a 0.2--12 keV flux of 5$\times$10$^{-12}$ erg cm$^{-2}$ 
s$^{-1}$, and a still unnamed source with a 0.2--12 keV flux of 
1.5$\times$10$^{-12}$ erg cm$^{-2}$ s$^{-1}$. Thus, a deeper 
multiwavelength effort is needed to shed light on the actual nature of 
HESS J1632$-$478 and on its connection with any of these three sources 
detected with {\it XMM-Newton}.
One should also note that HESS J1632$-$478 and HESS J1634$-$472
are extended and persistent, while all of the X--ray objects mentioned 
above are pointlike objects; moreover, 4U 1630$-$47 and IGR 
J16358$-$4726 are transients. So one can, at the very least, state that 
none of these two latter X--ray objects is the likely counterpart of the 
source HESS J1634$-$472.

\begin{figure}
%\begin{center}
\mbox{\psfig{figure=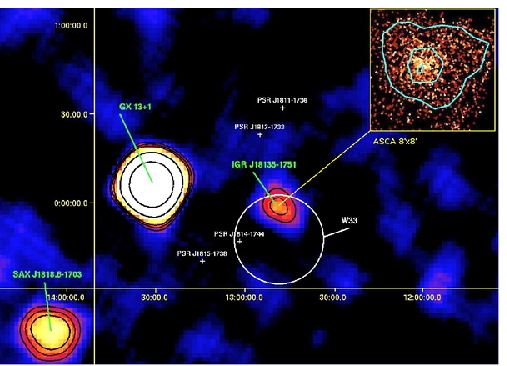,height=57mm,angle=0.0}}
%\vspace{1cm}
\psfig{figure=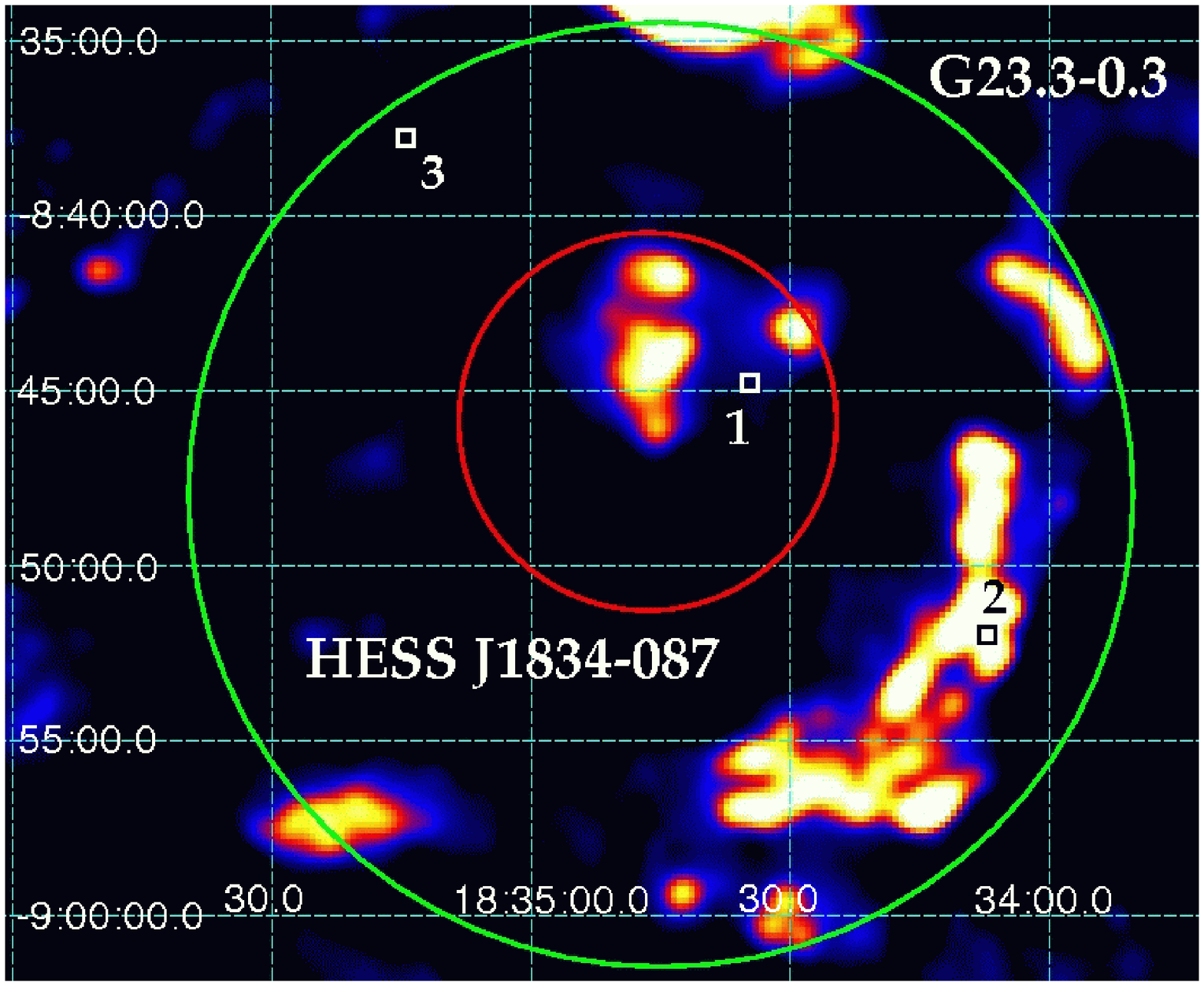,height=57mm,angle=0.0}
\caption{{\it Left panel}: IBIS/ISGRI 20-40 keV image
showing the location of HESS J1813$-$178 and its surroundings.
The extension of the TeV source is contained within the
contour of the {\it INTEGRAL} source IGR J18135$-$1751. Also shown are 
the location of SNR W33 (G12.82$-$0.02) and of the 4 nearest radio pulsars. 
The extension (15$'$ radius) of W33 is also indicated by the white circle.
The {\it ASCA} image at the {\it HESS} position is shown as an inset 
on the top right side of the figure; the box covers an 8$'$$\times$8$'$
region centered on the {\it ASCA} source position. Coordinates are
displayed in the Galactic system.
For details see Ubertini et al. (2005). {\it Right panel}: radio image at 
20 cm of the region surrounding HESS J1834$-$087. The larger circle 
(green, 13$'$ radius) describes the position and extension of SNR 
G23.3$-$0.3 as given in Green (2004). The smaller circle (red, 
3$'$ radius) represents instead the extension of the TeV source. 
The positions of the XRT sources are given by boxes. For details see Landi 
et al. (2006).}
%\end{center}
\end{figure}

{\bf HESS J1640$-$465}. This {\it HESS} source is positionally associated 
with the broken shell SNR 338.3-0.0 (Fig. 4, right panel), detected in 
X--rays by {\it ASCA} as AX J1640$-$4632 (Sugizaki et al. 2001). This 
X--ray source is well detected also by {\it Swift}/XRT at the centre of 
the {\it HESS} error circle, with spectrum and flux compatible with those 
measured with {\it ASCA} (Landi et al. 2006). Again according to these 
authors, the chance positional probability of finding a soft X--ray source 
with the flux of AX J164042$-$4632 (7.2$\times$10$^{-13}$ erg cm$^{-2}$ 
s$^{-1}$ in the 2--10 keV band) in the {\it HESS} error box is as low as 
1\%, suggesting that the two sources are the same. This evidence suggests 
that HESS J1640$-$465 is a SNR, although the lack of evident diffuse 
emission, together with the central location of the X--ray source within 
the SNR, do not allow one to rule out that this is actually a PWN. Recent 
{\it XMM-Newton} data (Funk et al. 2007) show extended emission from the 
{\it ASCA} source, and strengthen the SNR interpretation for the TeV 
emission observed by {\it HESS}. However, the {\it XMM-Newton} position of 
the source is not consistent with that of XRT. Moreover, in the {\it 
XMM-Newton} error circle one can find two optical sources, with magnitudes 
$R\sim$ 15.4 and $R\sim$ 17, respectively, whereas the XRT error box does 
not contain any optical or NIR object. Thus, optical spectroscopy of the 
sources within the {\it XMM-Newton} error circle is desirable to shed 
light on this positional mismatch and on their association with the X--ray 
and TeV emissions.

{\bf HESS J1804$-$216}. This source was observed with several soft X--ray 
satellites. {\it Swift}/XRT detected 3 objects within the TeV error circle 
of this object: two are likely stars, and the third is close to a radio 
complex of unknown origin (Landi et al. 2006). Subsequent observations by 
{\it Suzaku} detected 2 non-variable objects (S1 and S2, the latter 
coinciding with the third {\it Swift} source mentioned above), with hard 
and absorbed X--ray spectra (Bamba et al. 2007). More recently, 
observations with {\it Chandra} (Kargaltsev et al. 2007) showed that 
source S2 is compact, while S1 is extended. No reliable optical/NIR 
counterpart was found for both sources (Kargaltsev et al. 2007). All these 
observations still do not allow to understand what these two X--ray 
objects are, as well as what their connection (if any) with the TeV source 
is. Thus, the study of their X--ray variability is needed to shed light on 
their nature.

{\bf HESS J1813$-$178}. Ubertini et al. (2005) reported on a followup on 
this TeV source made with {\it INTEGRAL} and using {\it ASCA} archival 
data (Fig. 5, left panel). They found that the spectrum of the X--ray 
source positionally consistent with the TeV emission is modeled with an 
absorbed power law with photon index $\Gamma \sim$ 1.8, and hypothesized 
that this source is a PWN able to emit TeV photons. Radio data acquired 
with the VLA confirmed that the shell-type SNR G12.82$-$0.02 is 
positionally coincident with HESS J1813$-$178 (Brogan et al. 2005). More 
recently, {\it Chandra} data suggest the presence of a young X--ray pulsar 
within the SNR (Helfand et al. 2007). Therefore, this SNR/PWN is the 
likely counterpart of the {\it HESS} source.

{\bf HESS J1825$-$137}. The X--ray observations performed with {\it 
INTEGRAL} (A. Malizia, priv. comm.) and {\it XMM-Newton} (Gaensler et al. 
2003) allow associating this {\it HESS} source with the PWN of the X--ray 
pulsar PSR B1823$-$13. Actually, as in the case of HESS J1616$-$508, the 
pulsar is slightly outside the TeV error ellipse; however, the X--ray 
emitting nebula associated with the pulsar falls within the {\it HESS} 
error box. This supports the fact that the PWN powered by PSR B1823$-$13 
is responsible for the TeV emission from HESS J1825$-$137.

\begin{figure}
%\begin{center}
\mbox{\psfig{figure=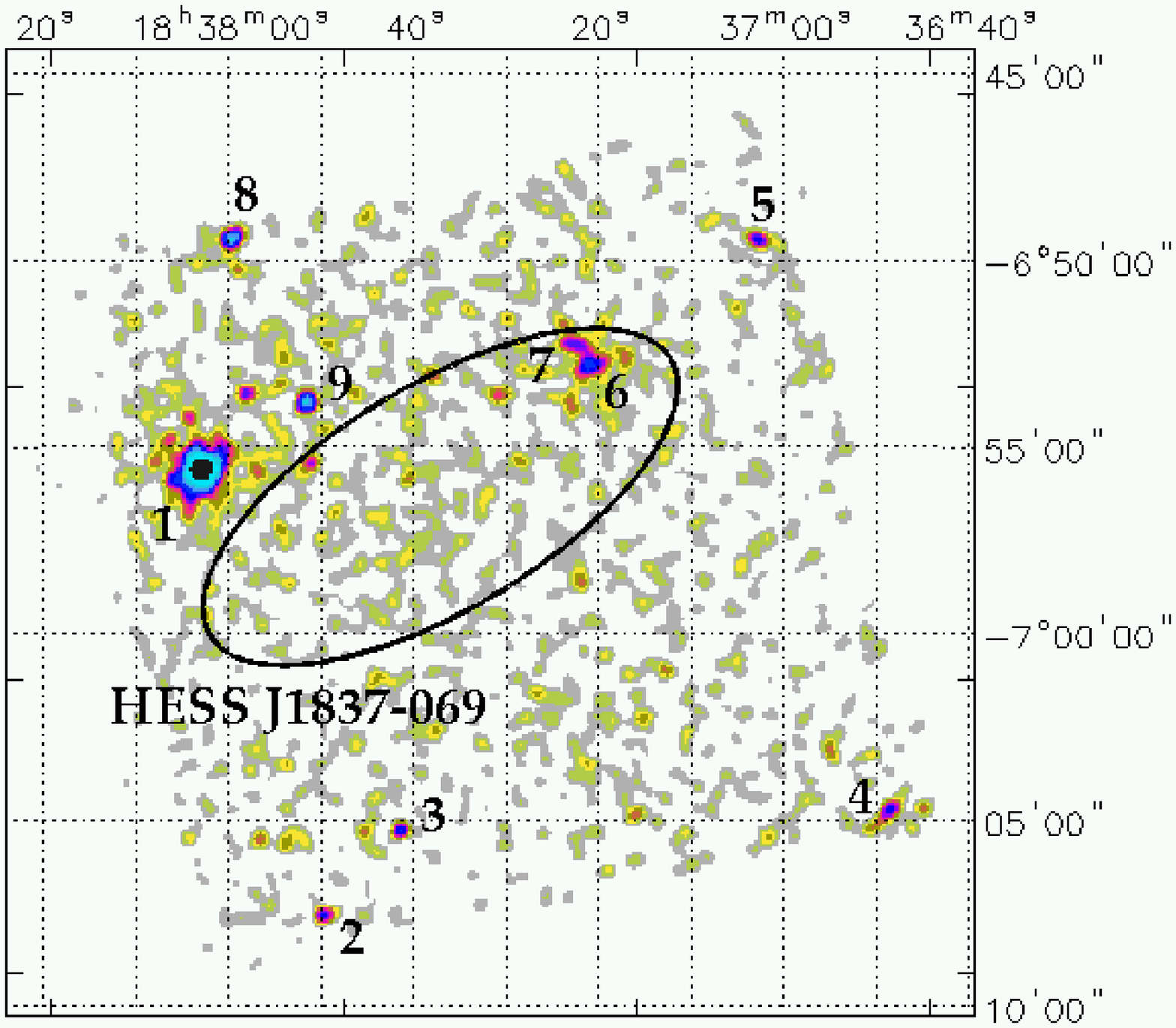,width=75mm,angle=0.0}}
%\vspace{1cm}
\psfig{figure=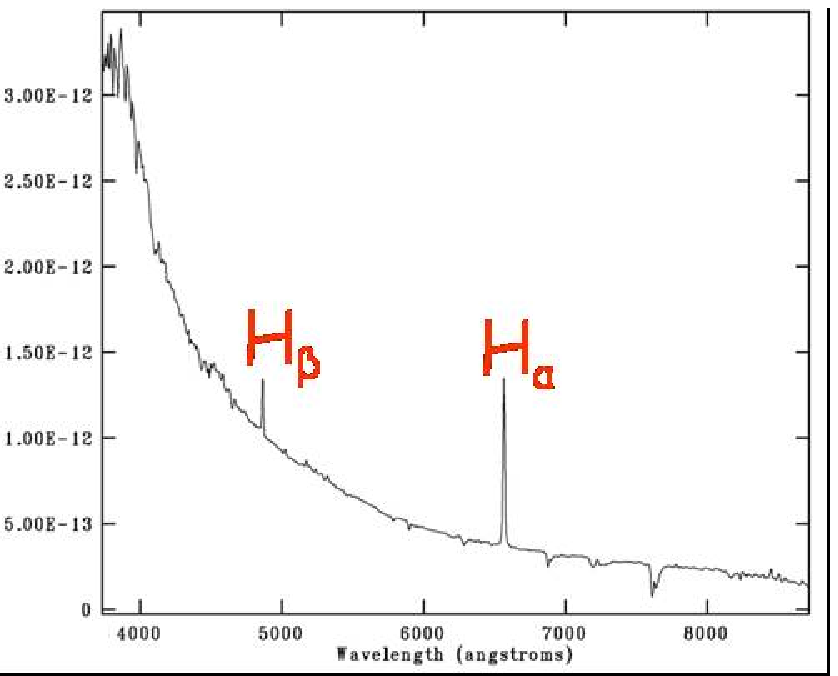,width=80mm,angle=0.0}
\caption{{\it Left panel}: XRT 0.3--10 keV image of the region surrounding 
HESS J1837$-$069. The ellipse (with axes of size 14$'$ and 6$'$) 
represents the extension of the TeV source.
For details see Landi et al. (2006). {\it Right panel}: optical spectrum
of the emission-line star HD 259440 acquired in Loiano on April 2007.
Balmer H$_\alpha$ and H$_\beta$ emissions, superimposed on a very blue 
continuum, are apparent. On the $y$-axis, the fluxes are expressed in
erg cm$^{-2}$ s$^{-1}$ \AA$^{-1}$.}
%\end{center}
\end{figure}

{\bf HESS J1834$-$087}. The positional coincidence of this TeV source with 
the central hot spot of the broken shell SNR G23.3$-$0.3 (Fig. 5, right 
panel) suggested an association between the two sources. {\it Swift}/XRT 
found only one source inside the TeV error box, at a 2--10 keV band flux 
of 2.6$\times$10$^{-13}$ erg cm$^{-2}$ s$^{-1}$, implying a chance 
positional probability of about 40\% (Landi et al. 2006). Thus, the two 
sources are probably not associated. It should be however noted that a 
reddened object (with magnitudes $R\sim$ 18 and $K \sim$ 13) is found at 
the border of the XRT error circle. So, albeit this TeV source may be 
associated with a SNR, more multiwavelength followup observations are 
needed to confirm this connection and to understand whether the emission 
is rather produced by a PWN.

{\bf HESS J1837$-$069}. This source is positionally associated with the 
SNR G25.5+0.0 complex, detected in X--rays by {\it ASCA} (Bamba et al. 
2003), {\it BeppoSAX} and {\it INTEGRAL} (Malizia et al. 2005). 
Observations made with {\it Swift}/XRT detected 9 sources in or around the 
{\it HESS} error ellipse for this source (Fig. 6, left panel; Landi et al. 
2006). Of these, only two (at 2--10 keV fluxes of 4$\times$10$^{-13}$ erg 
cm$^{-2}$ s$^{-1}$ and 2$\times$10$^{-13}$ erg cm$^{-2}$ s$^{-1}$) are 
within this ellipse, while the brightest one (with 2--10 flux of 
1.1$\times$10$^{-11}$ erg cm$^{-2}$ s$^{-1}$) lies slightly outside it. 
This latter source is associated with a very reddened optical object (with 
magnitudes $R\sim$ 17.5 and $K\sim$ 10.5), while the other two sources 
have no optical counterpart but are associated with relatively bright NIR 
objects (both have $K\sim$ 13). Thus, further observations, especially at 
optical and NIR wavebands, are needed to understand the nature of these 
X--ray sources and to see whether one of them is associated to the TeV 
emitter.

{\bf HESS J0632+058}. Although this TeV source is not in the sample of 
Aharonian et al. (2006) -- indeed, it was dicovered later (Aharonian et 
al. 2007) -- I included it in this review because it is located as well 
along the Galactic Plane and it has a similar spectral shape at TeV 
energies when compared to the sources of the aforementioned sample. HESS 
J0632+058 lies close to the rim of the Monoceros SNR, and its position 
suggests an association with a faint {\it ROSAT} soft X--ray source, with 
an unidentified EGRET high-energy source, and/or with the early-type, 
emission-line star HD 259440 (Aharonian et al. 2007). Recent optical 
spectroscopy of this star, acquired in Loiano on April 2007 (Fig. 6, right 
panel), indeed shows Balmer H$_\alpha$ and H$_\beta$ emissions 
superimposed on a very blue continuum. If the optical and the TeV source 
are associated, this object may be a gamma-ray microquasar similar to the 
HMXBs LS 5039 (Aharonian et al. 2005b) and Cyg X-1 (e.g., Romero et al. 
2002). However, accurate pointed X--ray and radio observations are needed 
to confirm (or disprove) this hypothesis.

To conclude this Section, I would like to stress the following: as 
{\it INTEGRAL} preferentially detects PWNe rather than SNRs (L. Bassani,
priv. comm.; Bird et al. 2007), one may suggest that the {\it HESS} 
sources detected with {\it INTEGRAL}, i.e. HESS J1616$-$508, HESS 
J1813$-$178 and HESS J1837$-$069, and which are likely associated with the 
residual of a Supernova explosion, are possibly PWNe.

Of course, a long work of multiwavelength followup still needs to be done
to conclusively determine the nature of these TeV sources; however, the
present evidence seems to suggest that many of them are possibly connected
with the aftermath of Galactic Supernovae, or with HMXBs showing 
collimated emission (i.e. microquasars).

%% Authors can use \cite, \citep and \citet for citation.
%% You may also give a citation as 'Michel et al. 1992', and use Table~1 or Fig.~1
%% and so forth. Using \ref and \label for cross-references of Tables/Figures is
%% a good way in adjusting/adding/removing text, tables or figures.

\section{Conclusions}

In summary, the {\it INTEGRAL} and {\it HESS} observatories allowed 
opening wide two major windows of the electromagnetic spectrum on our 
Galaxy and beyond; their data, complemented with multiwavelength 
spectroscopic and imaging followup from radio to X--rays, are fundamental 
for the identification and the study of the high-energy sources newly 
discovered by these facilities. And, as more and more hard X--ray and TeV 
data are accumulated, we can get an increasingly better insight on the 
single sources emitting at such high energies, as well as on their 
classes.

\begin{acknowledgements}
I thank the organizers of the workshop, Franco Giovannelli and Lola 
Sabau-Graziati, for having given me the opportunity to present this 
review, and the LOC for the warm hospitality and the pleasant stay in 
Vulcano.
I also thank the referee, Sylvain Chaty, for several useful remarks which 
helped me to improve this written contribution.
\end{acknowledgements}

\bigskip
\bigskip
\noindent {\bf DISCUSSION}

\bigskip
\noindent {\bf SYLVAIN CHATY:} In your sample of optically identified {\it
INTEGRAL} sources, which is the percentage of Be HMXBs compared to that of
HMXBs hosting a blue supergiant? Do you have information on the absorption
towards these sources?

\bigskip
\noindent {\bf NICOLA MASETTI:} The HMXBs with supergiant companion are at 
least 4 out of the 11 in our sample. However they may be more, as in some 
cases we could not firmly identify the luminosity class of the companion 
star. Concerning the absorption, we find that 6 cases out of 11 are 
heavily reddened; however, as they lie along the Galactic Plane, the 
observed reddening may be interstellar and not local to the system.

\bigskip
\noindent {\bf SYLVAIN CHATY's comment:} we acquired a NIR spectrum of IGR 
J16320$-$4751, which you possibly associate with the {\it HESS} source HESS 
J1632$-$478, and it appears to be a HMXB hosting a blue supergiant star.

\bigskip
\noindent {\bf ARNON DAR:} Can you give the linear size of the {\it HESS} error 
box radius if placed at the distance of the Galactic Center?

\bigskip
\noindent {\bf NICOLA MASETTI:} assuming a (typical) radius of 3$'$ one gets a linear 
size of about 7 pc. This of course means that we cannot discount the possibility
that some of the X--ray counterparts proposed for the HESS sources are chance
coincidences, as already stressed by Landi et al. (2006).

\bigskip
\noindent {\bf STEFANO COVINO:} Which is the percentage of intrinsically 
obscured AGNs in your sample? 

\bigskip
\noindent {\bf NICOLA MASETTI:} Up to now we found that, out of the 36 
objects that we identified as AGNs, 20 are Seyfert 2 galaxies. This would 
mean that 56\% of these objects are intrinsically absorbed. However this 
percentage can be lowered if, among these cases, `'genuine' Seyfert 2s, 
i.e. without intrinsic nuclear absorption, are found (see the contribution 
of T. Boller to these proceedings).

\bigskip
\noindent {\bf FILIPPO FRONTERA:} Is it possible, on the basis of the 
{\it INTEGRAL}/IBIS X--ray spectra alone, to establish the nature of a given 
source?

\bigskip
\noindent {\bf NICOLA MASETTI:} The {\it INTEGRAL} spectra can of course 
give hints on the nature of the unidentified sources, but only optical or 
NIR followup spectroscopy can definitely pinpoint their actual nature.

\label{lastpage}


\begin{thebibliography}{99}
%% you can type \apj for ApJ, \aap for A&A, \apss for Ap&SS, etc. Please consult
%% the macro chjaa.cls. You can also find them in aasguide.tex (AASTeX for ApJ, AJ, PASP)
%% Please follow the format of ChJAA's reference list


\bibitem[]{}Adelman-McCarthy J., Agueros M.A., Allam S.S. et al., 2006, 
	ApJS, 162, 38

\bibitem[]{}Aharonian F., Akhperjanian A., Beilicke M. et al., 2002, A\&A, 
	393, L37

\bibitem[]{}Aharonian F., Akhperjanian A.G., Aye K.-M. et al., 2005a, 
	A\&A, 439, 1013

\bibitem[]{}Aharonian F., Akhperjanian A.G., Aye K.-M. et al., 2005b, 
	Science, 309, 746 

\bibitem[]{}Aharonian F., Akhperjanian A.G., Bazer-Bachi A.R. et al., 2006, 
	ApJ, 636, 777

\bibitem[]{}Aharonian F., Akhperjanian A.G., Bazer-Bachi A.R. et al., 2007,
	A\&A, 469, L1

\bibitem[]{}Bamba A., Ueno M., Koyama K., Yamauchi S., 2003, ApJ, 589, 253

\bibitem[]{}Bamba A., Koyama K., Hiraga J.S. et al., 2007, PASJ, 59S, 209

\bibitem[]{}Barlow E.J., Knigge C., Bird A.J. et al., 2006, MNRAS, 372, 224 

\bibitem[]{}Bird A.J., Malizia A., Bazzano A. et al., 2007, ApJS, 170, 175

\bibitem[]{}Bodaghee A., Courvoisier T.J.-L., Rodriguez J. et al., 2007, 
	A\&A, 467, 585

\bibitem[]{}Brogan, C.L., Gaensler, B.M., Gelfand, J.D. et al., 2005, 
	ApJ, 629, L105

\bibitem[]{}Chakrabatry D., Roche P., 1997, ApJ, 489, 254

\bibitem[]{}de Martino D., Matt G., Belloni T., Haberl F., Mukai K., 2004,
	A\&A, 415, 1009

\bibitem[]{}Filliatre P., Chaty S., 2004, ApJ, 616, 469

\bibitem[]{}Funk S., Hinton J.A., P\"uhlhofer G. et al., 2007, ApJ, 662, 517

\bibitem[]{}Gaensler B.M., Schulz N.S., Kaspi V.M., Pivovaroff M.J., 
	Becker W.E., 2003, ApJ, 588, 441

\bibitem[]{}Garcia M.R., Baliunas S.L., Doxsey R. et al., 1983, ApJ, 267, 291

\bibitem[]{}Green D.A., 2004, BASI, 32, 335

\bibitem[]{}Helfand D.J., Gotthelf E.V., Halpern J.P. et al., 2007, ApJ, 
	in press {\tt [arXiv:0705.0065]}

\bibitem[]{}Hinton J., 2004, New Astron. Rev., 48, 331

\bibitem[]{}Jones D.H., Saunders W., Colless M. et al., 2004, MNRAS, 355, 747

\bibitem[]{}Kaplan D.L., Levine A.M., Chakrabarty D. et al., 2007, ApJ, 
	661, 437

\bibitem[]{}Kargaltsev O., Pavlov G.G., Garmire G.P., 2007, ApJ, submitted
	{\tt [astro-ph/0701069]}

\bibitem[]{}Kennea J.A., Campana S., 2006, ATel 818

\bibitem[]{}Krivonos R., Revnivtsev M., Lutovinov A. et al., 2007, A\&A, submitted
	{\tt [astro-ph/0701836]}

\bibitem[]{}Landi R., Bassani L., Malizia A. et al., 2006, ApJ, 651, 190

\bibitem[]{}Landi R., Masetti N., Bassani L. et al., 2007a, ATel 1047

\bibitem[]{}Landi R., De Rosa A., Dean A.J. et al., 2007b, MNRAS, in press
	{\tt [arXiv:0707.0832]}

\bibitem[]{}Leyder J.-C., Walter R., Lazos M., Masetti N., Produit N., 2007, 
	A\&A, 465, L35 

\bibitem[]{}Malizia A., Bassani L., Stephen J.B. et al., 2005, ApJ, 630, L157

\bibitem[]{}Masetti N., 2006, ChJAA Suppl., 6, 143

\bibitem[]{}Masetti N., Dal Fiume D., Cusumano G. et al., 2002, A\&A, 382, 
	104

\bibitem[]{}Masetti N., Palazzi E., Bassani L., Malizia A., Stephen J.B., 2004, 
	A\&A, 426, L41

\bibitem[]{}Masetti N., Bassani L., Bird A.J., Bazzano A., 2005, ATel 528

\bibitem[]{}Masetti N., Mason E., Bassani L. et al., 2006a, A\&A, 448, 547

\bibitem[]{}Masetti N., Pretorius M.L., Palazzi E. et al., 2006b, A\&A, 449, 1139

\bibitem[]{}Masetti N., Bassani L., Bazzano A. et al., 2006c, A\&A, 455, 11

\bibitem[]{}Masetti N., Morelli L., Palazzi E. et al., 2006d, A\&A, 459, 21

\bibitem[]{}Masetti N., Bassani L., Bazzano A. et al., 2006e, ATel 815

\bibitem[]{}Masetti N., Bassani L., Dean A.J., Ubertini P., Walter R., 2006f, ATel 715

\bibitem[]{}Masetti N., Bassani L., Malizia A., Bird A.J., Ubertini P., 2006g, ATel 941

\bibitem[]{}Masetti N., Malizia A., Dean A.J., Bazzano A., Walter R., 
	2006h, ATel 957

\bibitem[]{}Masetti N., Orlandini M., Palazzi E., Amati L., Frontera F., 
	2006i, A\&A, 453, 295

\bibitem[]{}Masetti N., Morelli L., Cellone S.A. et al., 2007a, ATel 1033

\bibitem[]{}Masetti N., Cellone S.A., Landi R. et al., 2007b, ATel 1034

\bibitem[]{}Masetti N., Landi R., Pretorius M.L. et al., 2007c, A\&A, 470, 
	331

\bibitem[]{}Masetti N., Rigon E., Maiorano E. et al., 2007d, A\&A, 464, 277

\bibitem[]{}Matsumoto H., Ueno M., Bamba A. et al., 2007, PASJ, 59S, 199

\bibitem[]{}Paizis A., Nowak M.A., Chaty S. et al., 2007, ApJ, 657, L109

\bibitem[]{}Piatti A.E., Clari\'a J.J., Bica E., 2000, A\&A, 360, 529

\bibitem[]{}Romero G.E., Kaufman Bernad\'o M.M., Mirabel I.F., 2002, A\&A, 
	393, L61

\bibitem[]{}Stephen J.B., Bassani, Malizia A. et al., 2006, A\&A, 445, 869

\bibitem[]{}Sugizaki M., Mitsuda K., Kaneda H. et al., 2001, ApJS, 134, 77

\bibitem[]{}Torres M.A.P., Steeghs D., Jonker P.G., Burns C.R., Freedman 
	W.L., 2006, ATel 909

\bibitem[]{}Torii K., Kinugasa K., Toneri T. et al., 1998, ApJ, 404, L207
 
\bibitem[]{}Tueller J., Barthelmy S., Burrows D. et al., 2005a, ATel 669

\bibitem[]{}Tueller J., Gehrels N., Mushotzky R.F. et al., 2005b, ATel 591

\bibitem[]{}Ubertini P., Bassani L., Malizia A. et al., 2005, ApJ, 629, L109

\bibitem[]{} Winkler C., Courvoisier T.J.-L., Di Cocco G. et al., 2003,
        A\&A, 411, L1


\end{thebibliography}
\end{document}